\documentclass[english,letterpaper,twocolumn,showpacs,pra,aps]{revtex4}
\usepackage{graphicx}
\usepackage{amssymb}

\makeatletter

\large  

\input epsf

\makeatother

\usepackage{babel}
\makeatother
\begin{document}

\title{Geometrical interpretation of the Casimir effect}

\author{Eugene B. Kolomeisky$^{1}$ and Joseph P. Straley,$^{2}$}

\affiliation{$^{1}$Department of Physics, University of Virginia, P. O. Box 400714,
Charlottesville, Virginia 22904-4714, USA\\
$^{{2}}$Department of Physics and Astronomy, University of Kentucky,
Lexington, Kentucky 40506-0055, USA}

\begin{abstract}
Casimir forces are a manifestation of the change in the zero-point energy of the vacuum
caused by the insertion of boundaries.  We show how the Casimir force 
can be efficiently computed by consideration of the  
vacuum fluctuations that are \textit{suppressed} by the boundaries,    
and
rederive the scalar Casimir effects for a 
series of the Dirichlet geometries.  For the planar case a finite universal force 
is automatically found.  Consistent with other calculations of the effect, for curved geometries divergent (non-universal) 
expressions are encountered.  They are interpreted geometrically following Candelas and Deutsch (1979) as largely due to the divergent self-energy of the boundary contributing to the force.  
This viewpoint is supported by explicit calculations for a wedge-circular arc geometry in two dimensions where non-universal and universal contributions into the effect can be unambiguously separated.  We also give a heuristic derivation of the purely geometrical expression (Sen, 1981) for the non-universal piece of the  Casimir energy due to an arbitrary  smooth two-dimensional Dirichlet boundary of a compact region.  
\end{abstract}


\maketitle

\section{Introduction}

Casimir interactions are due to the macroscopic response of the physical vacuum to the introduction of boundaries.  They were first derived as an attractive force between perfectly conductive parallel plates induced by the zero-point motion of 
the electromagnetic field \cite{Casimir}  There is convincing experimental evidence for the reality of these forces \cite{experiment} and a vast body of literature dedicated to various aspects of the phenomenon \cite{CasReviews}.  

The  Casimir interaction $\mathcal{E}$ is the difference between the vacuum 
energy of the system constrained by the boundaries and 
that of free space.  Since boundaries made of real materials are transparent to sufficiently high-energy modes \cite{Casimir}, the high energy spectrum is unaffected by the geometry of the system, and only a finite range of the spectrum need be considered.  However, 
in the theoretical treatments of this effect the vacuum energies are usually calculated from an effective low-energy harmonic field 
theory (such as quantum electrodynamics in the case of the electromagnetic Casimir effect), so that
they are approximated by the sum of zero-point energies of a collection of simple harmonic 
oscillators with a spectrum $\omega = c |\textbf{k}|$ (where $c$ is the speed of 
light).  In this model, the dispersion relation holds for arbitrarily large wave vectors $\textbf{k}$; both the 
"constrained" and "free" vacuum energy densities are ultraviolet divergent; and 
the Casimir interaction is the difference between two infinite quantities.  
The theory resolves this problem by a soft-cutoff modification of the large-\textbf{k} part of the spectrum that leads to a finite vacuum energy.  The result is not very sensitive to the form of the cutoff, so that the Casimir interaction can then be extracted by taking the cutoff to infinity at the end of calculation.  Other approaches to the calculation that make use of analytic continuation \cite{zeta} and dimensional regularization \cite{dim} techniques give the same answer, thus adding to the credibility of the result.

As already 
noted, the divergences are more mathematical artifacts than physical reality.
The important virtue of the model is that for many geometries a finite universal result depending only on $\hbar$, $c$, and macroscopic length scales is obtained without introducing a cutoff.   However, this is not always the case: specifically, the divergences occurring for spherical geometry in even space dimensions do not cancel  \cite{Bender,Milton}.  What this means physically represents an open problem;  it seems to imply that in two dimensions
a conducting ring placed in vacuum is unstable.       

The goal of this paper is two-fold:  first, we show how the Casimir effect can be efficiently computed by direct consideration of the fluctuation modes that are eliminated by the presence of the boundaries.   Technically this is accomplished by using a method closely related to the path-integral approach to the Casimir effect  \cite{pathintegral}.  

Second, we compute the Casimir effect in a  two-dimensional wedge-circular arc geometry.  The results of our analysis lend support to the idea of  Candelas and Deutsch \cite{CD} that the divergences encountered in the case of curved boundaries are of geometrical nature and mostly due to divergent self-energy of the boundary contributing to the Casimir force.  Since the geometry we study is a relative of the ring geometry in two dimensions,  our results also shed some light on the physical meaning of the divergent Casimir force exerted on the ring in two dimensions \cite {Bender,Milton}.  

The organization of this paper is as follows.  In Section I we introduce a method of computation of the Casimir effect which is based on consideration of the vacuum fluctuations eliminated by introduction of boundaries.  The efficiency of this technique is illustrated in Section II where the scalar Casimir energies and pressures are re-derived for a series of standard geometries:  planar geometry in an arbitrary dimension,  circular geometry in two dimensions, cylindrical and spherical geometries in three dimensions, and finally, spherical geometry in an arbitrary space dimension.   We observe that the Casimir pressures are always divergent in curved geometries.     Following Candelas and Deutsch \cite{CD} we pursue an interpretation of these divergences as non-universal contributions due to the geometry of the boundary.  This is made explicit in detailed calculations of Section III where the effect is computed for the wedge-circular arc geometry in two spatial dimensions with the wedge edges subject to the Dirichlet and periodic boundary conditions.  The latter includes the circular geometry in two dimensions as a special case.  Our main observation here is that non-universal contributions to the Casimir effect can indeed be attributed to the geometry of the boundary.  Moreover the non-universal pieces of the effect turn out to be insensitive to the change of the topology of the vacuum while the universal ones are.  This fact is used in Section IV to heuristically re-derive a purely geometrical formula \cite{Sen} for the non-universal piece of the Casimir energy of an arbitrary smooth Dirichlet boundary of a compact region.  We conclude (Section V) by summarizing our findings.

\section{Formalism}

In what follows we will be analyzing Casimir effects in a Gaussian field theory with the Euclidian action 
\begin{equation}
\label{action}
S[w] = {\frac{1}{2}}\int_{0}^{\hbar/T} d\tau d^{d}x \left (c^{-2}({\frac{\partial w}{\partial \tau}})^{2} +
(\nabla w)^{2}\right ),
\end{equation}
where $T$ is the temperature.  The real scalar $w$ is a function of the $d$-dimensional position vector $\textbf{r}$ and imaginary time $\tau$, and is periodic on the Matsubara circle, $w(\textbf{r},0) = w(\textbf{r}, \hbar/T)$ \cite{Matsubara}.  The action (\ref{action}) is applicable at energies low compared to some scale $\hbar \omega_{0}$;  the  frequency cutoff $\omega_{0}$ is provided by the properties of the material the boundary is made of.

The scalar field theory (\ref{action})  can be viewed as a toy version of electricity and magnetism.  The divergences encountered for curved geometries exist in both theories, so they are not due to specifics of electricity and magnetism.  Therefore we study the problem in the simpler setting of the scalar theory. 

The zero-point energy can be calculated by means of a functional integral that makes use of the correspondence between the Feynman path integral for the $d$-dimensional field theory and the partition function for a $d + 1$-dimensional classical statistical mechanics problem.   The path integral is
\begin{equation}
\label{Zdef}
Z_{w} = \int Dw(\textbf{r},\tau) \exp( - S[w]/\hbar)
\end{equation}
but it can be interpreted as the partition function for a classical statistical mechanics problem with the Hamiltonian $S$ at a fictitious temperature which is equal to Planck's constant \cite{Kogut}.  The zero-point energy corresponds to the "free energy" per unit "length" in the imaginary time direction, so that  $\mathcal{E}_{0} = -\hbar (\ln Z_{w})/(\hbar /T) = - T \ln Z_{w}$.  

Now assume that the vacuum is disturbed by the addition of sharp boundaries which constrain the field in some way and thus eliminate some degrees of freedom of the 
vacuum fluctuations.   The constrained field (which we will refer to as $v$) inherits 
any boundary conditions imposed on $w$ as well as new conditions on surfaces $D_{i}$ and $N_{j}$ of Dirichlet ($v|_{D_{i}} = 0$) or Neumann ($\partial v/\partial n|_{N_{j}} = 0$) type, respectively, where the subscripts $i$ and $j$ label the boundaries and $\partial v/\partial n$ is the normal derivative. We will write the difference between the original and constrained fields in the form $w(\textbf{r},t) = v(\textbf{r},t) + u(\textbf{r},t)$, where $u(\textbf{r},t)$ can be chosen to satisfy the $d+1$-dimensional equation
\begin{equation}
\label{gbvproblem}
(\frac{\partial^{2}}{c^{2}\partial\tau^{2}} + \triangle) u = 0, ~u|_{D_{i}} = f_{i}(\textbf{r},\tau),\frac{\partial u}{\partial n}|_{N_{j}} = g_{j}(\textbf{r}, \tau)
\end{equation} 
where $f_{i}$ and $g_{j}$ are functions defined on the boundaries and determined by the boundary values of $w$; they play the role of dynamical variables of our approach.  The reason for defining $u$ this way is that it eliminates the cross term in the action, so that $S[w] = S[v] + S[u]$.  Then the functional integral factors into integrations over $v$ and $u$, so that $Z_{w} = Z_{v} Z_{u}$.  Since the zero-point energy for the unconstrained system is determined by $Z_{w}$  
and the zero-point energy for the constrained system is determined by $Z_{v}$, the Casimir energy is given by $\mathcal{E}= T \ln Z_{u}$.  This result allows us to calculate directly the change in energy due to the fluctuation modes that have been eliminated.  

A simplification is achieved by expanding all the dynamical variables of the problem into a Fourier series in the imaginary time domain; for example $u(\textbf{r}, \tau) = \sum_{\omega} u_{\omega}(\textbf{r}) \exp i
\omega \tau$
where the Fourier coefficients $u_{\omega}(\textbf{r})$ are solutions to the boundary-value problem for the Helmholtz equation
\begin{equation}
\label{Helmholtz}
(\triangle - \frac{\omega^{2}}{c^{2}})u_{\omega} = 0,~ u_{\omega}|_{D_{i}} = f_{\omega,i}(\textbf{r}),~\frac{\partial u_{\omega}}{\partial n}|_{N_{j}} = g_{\omega,j}(\textbf{r})
\end{equation}  

The calculation of the action $S[u]$ is further  simplified when the identity $(\nabla u)^{2} = div(u\nabla u) - u\triangle u$ is substituted into Eq.(\ref{action}).  Then the integral of $div(u\nabla u)$ over $d^{d}x$ transforms into a sum of surface integrals.  The remaining integral over $d\tau$ vanishes due to the relation $\triangle u = - \partial^{2} u/c^{2}\partial \tau^{2}$ and the condition of periodicity, $u(\textbf{r},0) = u(\textbf{r}, \hbar/T)$.  As a result we find
\begin{eqnarray}
\label{actionsurfaceintegral}
S[u]& =& \frac{1}{2} \int_{0}^{\hbar/T} d \tau \sum_{i}\int[u\nabla u]_{i}d\textbf{s}_{i} \nonumber\\& = & \frac{\hbar}{2T}\sum_{\omega,i}\int[u_{\omega}\nabla u_{-\omega}]_{i} d\textbf{s}_{i}
\end{eqnarray}
Here $[\psi]_{i}$ stands for the discontinuity of $\psi$ across the $i$-th boundary, and the summation is performed over all the boundaries.  

Although our approach is applicable to an arbitrary number of the Dirichlet and/or Neumann boundaries, in all the cases considered in this paper only one Dirichlet boundary contributes into the integral (\ref{actionsurfaceintegral}).  Since we are dealing with a harmonic field theory, the solution to the boundary-value problem (\ref{Helmholtz}) is linear in the surface field $f_{\omega}$ and thus the action $S_{E}$ is a quadratic diagonal form of $f_{\omega}$:
\begin{equation}
\label{form}
S_{E} = \frac{\hbar}{2T} \sum_{\omega,\nu} \frac{|f_{\omega \nu}|^{2}}{\lambda_{\nu}(|\omega|/c)}
\end{equation}
where the subscript $\nu$ (representing one or more indices) labels the normal modes of the field $u$ that have been eliminated by the boundary in question, and $\lambda_{\nu}(|\omega|/c) > 0$ is a set of geometry-dependent coefficients.  Then according to the prescription outlined above, the Casimir energy is given by
\begin{eqnarray}
\label{Casenergygeneral}
\mathcal{E}&= &\frac{T}{2} \sum_{\omega, \nu} \ln\frac{2\pi T \lambda_{\nu}(|\omega|/c)}{\hbar}\nonumber\\ &\rightarrow& \frac{\hbar c}{2\pi} \sum_{\nu}\int_{0}^{\infty}d\kappa\ln \lambda_{\nu}(\kappa)\nonumber\\
&\rightarrow& - \frac{\hbar c}{2\pi}\sum_{\nu}\int_{0}^{\infty}\kappa d\kappa F(\frac{\kappa c}{\omega_{0}})\frac{d}{d\kappa}\ln \lambda_{\nu}(\kappa)
\end{eqnarray}
where in the second step we took the zero-temperature limit according to the rule $\sum_{\omega} \rightarrow (\hbar/T) \int d\omega/2\pi$ and dropped  contributions that are independent of geometry.  In the third representation, a monotonic cutoff function $F(y)$ rapidly decaying for $y > 1$ and satisfying $F(0) = 1$ is made explicit;   the cutoff prescription, $\hbar c\kappa/2 = \hbar \omega/2 \rightarrow (\hbar\omega/2)F(\omega/\omega_{0})$,  reflects penetrability of the boundary to high-energy modes\cite{Sen}.  In most of the formulas below the cutoff function is suppressed and invoked only as needed;  analysis of such cases is conducted for an arbitrary cutoff function.  We note that the last representation of (\ref{Casenergygeneral}) hints at a relationship of our approach to a contour integral method of calculation of the Casimir energy \cite{Milton}. 

\subsection{Planar geometry}

Consider three Dirichlet planes at $z = 0$, $z = a$, and $z = L$, where $z$ is one of the axes of the $d$-dimensional rectangular coordinate system and $0 < a < L$.   We are interested in the Casimir pressure exerted on the middle partition at $z = a$.  The outer boundaries are fixed in place, so that there is no need to look beyond them.  Since the space is uniform relative to translations parallel to the boundaries, the field $u_{\omega}(\textbf{r})$ is expanded into a Fourier series $u_{\omega}(\textbf{r}) = \sum_{\small\textbf{q}} u_{\omega \small\textbf{q}}(z) \exp i \textbf{q} \textbf{r}_{\perp}$ where $\textbf{r}_{\perp}$ is the position vector perpendicular to the $z$ axis.  Then the boundary-value problem (\ref{Helmholtz}) for the Fourier coefficients $u_{\omega \small\textbf{q}}(z)$ becomes
\begin{equation}
\label{bvplane}
(\frac{d^{2}}{dz^{2}} - q^{2} - \frac{\omega^{2}}{c^{2}})u_{\omega \small\textbf{q}} = 0, ~u_{\omega \small\textbf{q}}|_{0,L} = 0,~u_{\omega \small\textbf{q}}|_{a} = f_{\omega \small\textbf{q}}
\end{equation}
The particular solution to (\ref{bvplane}) is
\begin{eqnarray}
\label{planesolution}
u_{\omega \small\textbf{q}}(z)& = &f_{\omega \small\textbf{q}}\frac{\sinh (|\kappa| z)}{\sinh (|\kappa| a)}, ~~~~~~0 \leqslant z \leqslant a, \kappa^{2} = q^{2} + \frac{\omega^{2}}{c^{2}}\nonumber\\
u_{\omega \small\textbf{q}}(z)& = &f_{\omega \small\textbf{q}}\frac{\sinh (|\kappa| (L - z))}{\sinh (|\kappa| (L - a))}, ~a < z \leqslant L
\end{eqnarray}
Substituting this in Eq.(\ref{actionsurfaceintegral}) we see that only the partition at $z = a$ contributes to the action $S_{E}$ with the result 
\begin{equation}
\label{actionplanar}
S_{E} = \frac{\hbar \mathcal{A}}{2T} \sum_{\omega, \small\textbf{q}} |\kappa| (\coth (|\kappa| a) + \coth (|\kappa|(L - a)))|f_{\omega \small\textbf{q}}|^{2}
\end{equation}
where $\mathcal{A}$ is the macroscopic $(d - 1)$-dimensional area of the 
boundary.
Eq.(\ref{actionplanar}) has the required form Eq.(\ref{form}), with coefficients
that become large ($\lambda$ becomes small) for large $\kappa$, because 
for high frequencies or large transverse wavevector the disturbance
introduced by the fluctuating boundary condition is localized at the
boundary, to within a length that is proportional to $\lambda$ itself.
This will lead in what follows to a divergent surface energy.

The Casimir energy per unit area can be deduced from
Eq.(\ref{actionplanar})
according to the rule Eq.(\ref{Casenergygeneral}), with the result
\begin{eqnarray}
\label{Casimirenergyplanar}
&&\frac{\mathcal{E}}{\mathcal{A}} = -\frac{T}{2\mathcal{A}} \sum_{\omega, \small\textbf{q}} \ln \left (\frac{\coth (|\kappa| a) + \coth (|\kappa|(L - a))}{2\pi T/\hbar \mathcal{A} |\kappa|}\right )\nonumber\\
& \rightarrow & -\frac{\hbar}{2} \int \frac{d\omega d^{d - 1}q}{(2\pi)^{d}} \ln \left (\coth (|\kappa | a) + \coth (|\kappa |(L - a))\right )\nonumber\\
& = & - \frac{\hbar c K_{d}}{2} \int _{0}^{\infty}\kappa^{d-1} d\kappa \ln \left (\coth (\kappa a) + \coth (\kappa (L - a))\right )\nonumber\\
\end{eqnarray}
where in taking the macroscopic limit we used the rule $\sum_{\small\textbf{q}} \rightarrow \mathcal{A} \int d^{d-1}q/(2\pi)^{d-1}$.  Additionally, in going from the first to the second representation we took the $T = 0$ limit and dropped all $a$-independent contributions not influencing the pressure on the partition.  Hereafter all such contributions will be systematically dropped; however, we should note that in the present case part of what has been omitted is a divergent integral representing an infinite surface energy.  The parameter $K_{d}$ in the third representation is the surface area of a $d$-dimensional unit sphere, $S_{d}=2\pi^{d/2}/\Gamma(d/2)$, divided by $(2\pi)^{d}$.  The Casimir pressure on the boundary, $\mathcal{P} = - \partial (\mathcal{E}/\mathcal{A})/\partial a$ can be found in closed form
\begin{equation}
\label{planarpressure}
\mathcal{P} =  \frac{d \Gamma(\frac{d+1}{2}) \zeta(d + 1)} {(4\pi)^{\frac{d+1}{2}}}\hbar c\left (\frac{1}{(L - a)^{d + 1}} - \frac{1}{a^{d+1}}\right )
 \end{equation}   
 where $\Gamma(x)$ and $\zeta(x)$ are Euler's and Riemann's gamma and zeta functions, respectively \cite{AS}.  In arriving at (\ref{planarpressure}) we used the gamma function duplication formula, $\Gamma(z)\Gamma(z+\frac{1}{2})=2^{1-2z} \sqrt{\pi}\Gamma(2z)$ \cite{AS} and the value of the integral \cite{integral}
 \begin{equation}
 \label{integral}
 \int_{0}^{\infty}x^{d}(\coth x - 1)dx = 2^{-d}\Gamma(d + 1) \zeta(d + 1)
 \end{equation}

We see that the partition at $z = a$ is attracted to the closest outer boundary; specifically, in one dimension Eq.(\ref{planarpressure}) reduces to the well-known result \cite{Boyer}.  Since the outer boundaries impose the Dirichlet boundary conditions we can imagine joining them together.  Then Eq.(\ref{planarpressure}) describes Casimir interaction between two boundaries;  taking the $L \rightarrow \infty $ limit we then reproduce the result of Ambj\o rn and Wolfram \cite{dim}.

\subsection{Circular geometry}

Consider a Dirichlet circle of radius $a$ in two spatial dimensions.  The boundary-value problem (\ref{Helmholtz}) for this geometry becomes
\begin{equation}
\label{bvcircle}
\left (\frac{1}{\rho}\frac{\partial}{\partial \rho}(\rho \frac{\partial}{\partial \rho}) + \frac{1}{\rho^{2}}\frac{\partial^{2}}{\partial\varphi^{2}} - \frac{\omega^{2}}{c^{2}}\right )u_{\omega} = 0,u_{\omega}|_{a} = f_{\omega}(\varphi)
\end{equation}
where $\rho$ and $\varphi$ are the polar coordinates.  Seeking the particular solution in the form, $u_{\omega}(\rho, \varphi) = R_{\omega}(\rho)\exp i n\varphi$, where $n$ is an arbitrary integer, we find that the radial function $R_{\omega}(\rho)$ satisfies  the equation \begin{equation}
\label{Bessel}
\frac{d^{2}R_{\omega}}{d\rho^{2}} + \frac{1}{\rho}\frac{dR_{\omega}}{d\rho} - (\frac{\omega^{2}}{c^{2}} + \frac{n^{2}}{\rho^{2}})R_{\omega} = 0
\end{equation}
whose linearly-independent solutions are modified Bessel functions $I_{n}(|\omega|\rho/c)$ and $K_{n}(|\omega|\rho/c)$ \cite{AS}.  Thus the particular  solution to the boundary-value problem (\ref{bvcircle}) finite at $\rho = 0$ and decaying as $\rho \rightarrow \infty$  is
\begin{eqnarray}
\label{circle solution}
u_{\omega}& = &\sum_{n=-\infty}^{\infty}\frac{I_{n}(|\omega |\rho/c)}{I_{n}(|\omega |a/c)} f_{\omega n} \exp in\varphi, ~~~\rho \leqslant a \nonumber\\
u_{\omega}& = &\sum_{n=-\infty}^{\infty}\frac{K_{n}(|\omega |\rho/c)}{K_{n}(|\omega |a/c)} f_{\omega n} \exp in\varphi, ~~~\rho > a
\end{eqnarray}
At the circle $\rho = a$ both of these reduce to $f_{\omega}(\varphi) =  
\sum_{n
 = -\infty}^{\infty} f_{\omega n}\exp i n\varphi$.  Substituting the solution (\ref{circle solution}) in Eq.(\ref{actionsurfaceintegral})
and performing the angular integration we find
\begin{equation}
\label{circleaction}
S_{E} = \frac{a\pi \hbar}{cT} \sum_{\omega,n}
|\omega| \left (\frac {I_{n}^{'}(|\omega| a/c)}{I_{n}(|\omega| a/c)}
-\frac {K_{n}^{'}(|\omega| a/c)}{K_{n}(|\omega| a/c)}\right )
|f_{\omega n}|^{2}
\end{equation}
We observe that for large $n$ or large $\omega$ the prefactor in this sum is
large, because the functions
Eq.(\ref{circle solution})
are localized near the circle boundary.  This result can be simplified
by application of the Wronskian relationship
$ K_{n}(z)I_{n}^{'}(z) - I_{n}(z)K_{n}^{'}(z) = 1/z$ \cite{AS}.
The Casimir energy $\mathcal{E}$ and the pressure $\mathcal{P} = - (2\pi a)^{-1} \partial \mathcal{E}/\partial a$  are then implied by Eq.(\ref{Casenergygeneral})
\begin{equation}
\label{Casenergycircle}
\mathcal{E} = \frac{\hbar c}{2\pi}\sum_{n = -\infty}^{\infty} \int_{0}^{\infty} d\kappa \ln\left (I_{n}(\kappa a)K_{n}(\kappa a)\right )
\end{equation}
\begin{equation}
\label{Casforcecircle}
\mathcal{P} = -\frac{\hbar c}{4\pi^{2} a^{3}} \sum_{n = -\infty}^{\infty} \int_{0}^{\infty}xdx \frac{d}{dx}\ln(I_{n}(x)K_{n}(x))
\end{equation}
The results (\ref{Casenergycircle}) and (\ref{Casforcecircle}) are due to Sen \cite{Sen} (see also Refs.  \cite{Ng,Bender}).

\subsection{Cylindrical geometry}

The Casimir interaction for three-dimensional infinitely long Dirichlet cylinder of radius $a$ can be inferred from corresponding results for the circle.  First, expand the field $u_{\omega}(\textbf{r})$ into a Fourier series, $u_{\omega}(\rho, \varphi, z) = \sum_{q} u_{\omega q_{z}}(\rho, \varphi) \exp i q_{z}z$ where $\rho$, $\varphi$, and $z$ are cylindrical coordinates, and the $z$ axis coincides with that of the cylinder.  The Fourier coefficients $u_{\omega q_{z}}(\rho, \varphi)$ are solutions to the boundary value problem (\ref{bvcircle}) except that the role of $\omega^{2}/c^{2}$ is played by $q_{z}^{2} + \omega^{2}/c^{2}$.  Thus cylindrical analogs of Eqs.(\ref{Bessel})-(\ref{circleaction}) can be written down by replacing the summation index $\omega$ by the combination $(\omega , q_{z})$, and  $\omega^{2}/c^{2}$ by $\kappa^{2} = q_{z}^{2} + \omega^{2}/c^{2}$.  Moreover, an extra factor of the macroscopic cylinder length $\mathcal{L}$ appears in the counterpart to Eq.(\ref{circleaction}).  As a result the Casimir energy per unit length $\mathcal{E}/\mathcal{L}$ and the pressure on the boundary $\mathcal{P} = - (2\pi a\mathcal{L})^{-1}\partial \mathcal{E}/\partial a$ are given by  
\begin{equation}
\label{casenergycylinder}
\frac{\mathcal{E}}{\mathcal{L}} = \frac{\hbar c}{4\pi}\sum_{n = -\infty}^{\infty} \int_{0}^{\infty} \kappa d\kappa \ln\left (I_{n}(\kappa a)K_{n}(\kappa a)\right )
\end{equation}
\begin{equation}
\label{Caspressurecylinder}
\mathcal{P} = -\frac{\hbar c}{8\pi^{2} a^{4}} \sum_{n = -\infty}^{\infty} \int_{0}^{\infty}x^{2}dx \frac{d}{dx}\left(\ln(I_{n}(x)K_{n}(x))\right )
\end{equation}
respectively.  Again, these results are known \cite{Nesterenko}.

\subsection{Spherical geometry in three dimensions}

Consider a Dirichlet sphere of radius $a$ in three spatial dimensions.  The boundary-value problem (\ref{Helmholtz}) for this geometry becomes     
\begin{eqnarray}
\label{bvsphere}
\Big\{ \frac{1}{r^{2}}\frac{\partial}{\partial r}(r^{2}\frac{\partial}{\partial r}) &+& \frac{1}{r^{2}}\left (\frac{1}{\sin \theta}\frac{\partial}{\partial \theta}(\sin \theta\frac{\partial}{\partial \theta})+\frac{1}{\sin^{2}\theta}\frac{\partial}{\partial \varphi^{2}}\right ) \nonumber\\
&-&\frac{\omega^{2}}{c^{2}}\Big\} u_{\omega} = 0, ~~u_{\omega}|_{a} = f_{\omega}(\theta, \varphi)
\end{eqnarray}
where $r$, $\theta$ and $\varphi$ are spherical coordinates.  Seeking the particular solution in the form, $u_{\omega}(r, \theta, \varphi) = R_{\omega}(r) Y_{lm}(\theta, \varphi)$ where $Y_{lm}(\theta, \varphi)$ is a spherical function of order $l$, $m$ \cite{QM}, we find that the radial function $R_{\omega}(r)$ satisfies the equation
\begin{equation}
\label{spBessel}
\frac{d^{2}R_{\omega}}{dr^{2}} + \frac{2}{r}\frac{dR_{\omega}}{dr} - \left (\frac{\omega^{2}}{c^{2}} + \frac{l(l+1)}{r^{2}}\right )R_{\omega} = 0
\end{equation}
whose linearly-independent solutions are $I_{l + \frac{1}{2}}(|\omega|r/c)/\sqrt{r}$ and $K_{l + \frac{1}{2}}(|\omega|r/c)/\sqrt{r}$ \cite{QMnote}.  Thus the particular solution to the boundary-value problem (\ref{bvsphere}) finite at the origin $r = 0$ and decaying as $r \rightarrow \infty$ is 
\begin{eqnarray}
\label{sphere solution}
u_{\omega}& = &\sum_{l=0}^{\infty}\sum_{m=-l}^{l}\frac{\sqrt{a}I_{l+\frac{1}{2}}(|\omega |r/c)}{\sqrt{r}I_{l+\frac{1}{2}}(|\omega |a/c)} f_{\omega lm} Y_{lm}(\theta, \varphi), ~r \leqslant a \nonumber\\
u_{\omega}& = &\sum_{l=0}^{^{\infty}}\sum_{m=-l}^{l}\frac{\sqrt{a}K_{l+\frac{1}{2}}(|\omega |r/c)}{\sqrt{r}K_{l+\frac{1}{2}}(|\omega |a/c)} f_{\omega lm} Y_{lm}(\theta, \varphi), ~r > a \nonumber\\
\end{eqnarray}
At the sphere surface $r = a$ both of these reduce to $f_{\omega}(\theta, \varphi) = \sum_{l,m}f_{\omega lm}Y_{lm}(\theta, \varphi)$.  Substituting the solution (\ref{sphere solution}) in Eq.(\ref{actionsurfaceintegral}) and performing angular integration we find
\begin{equation}
\label{sphereaction}
S_{E} = \frac{\hbar a}{2T}\sum_{\omega,l,m}\frac{|f_{\omega lm}|^{2}}{I_{l+\frac{1}{2}}(|\omega|a/c)K_{l+\frac{1}{2}}(|\omega|a/c)}
\end{equation}
In arriving at (\ref{sphereaction}), similar to the circular case,  we used the Wronskian of the $I_{l+1/2}(z)$ and $K_{l+1/2}(z)$ pair as well as the property of orthogonality of the spherical functions \cite{QM}.  As a result the Casimir energy $\mathcal{E}$ and the pressure on the boundary $\mathcal{P} = -(4\pi a^{2})^{-1}\partial \mathcal{E}/\partial a$ are implied by Eq.(\ref{Casenergygeneral})
\begin{equation}
\label{Casenergysphere}
\mathcal{E} = \frac{\hbar c}{2\pi}\sum_{l=0}^{\infty}(2l+1)\int_{0}^{\infty}d\kappa
\ln \frac{I_{l+\frac{1}{2}}(\kappa a) K_{l+\frac{1}{2}}(\kappa a)}{\kappa a}
\end{equation} 
\begin{equation}
\label{Caspressuresphere}
\mathcal{P} = - \frac{\hbar c}{8\pi^{2}a^{4}}\sum_{l=0}^{\infty}(2l+1)\int_{0}^{\infty}xdx\frac{d}{dx}\ln \frac{I_{l+\frac{1}{2}}(x)K_{l+\frac{1}{2}}(x)}{x}
\end{equation}
These equations reproduce the results due to Bender and Milton  \cite{Bender} who also derived more general relationships for a Dirichlet sphere in space of arbitrary dimension. 

\subsection{Spherical geometry in arbitrary number of dimensions}

As a  last demonstration of the efficiency of our technique we derive the expressions for the Casimir energy and pressure for a Dirichlet sphere of radius $a$ in arbitrary space dimension.  Before proceeding we remind the reader \cite{Sommerfeld,Muller,Mikhlin} that in $d$ dimensions spherical functions can be arrived at by first enumerating all homogeneous linearly-independent harmonic polynomials of degree $l$, $P_{l,d}(\textbf{r})$.  After the transformation to spherical coordinates  $x_{1}=r\cos\varphi_{1}$, $x_{2}=r\sin\varphi_{1}\cos\varphi_{2}$, ..., $x_{d-1}=r\sin\varphi_{1}\sin\varphi_{2}...\sin\varphi_{d-2}\cos\varphi_{d-1}$, $x_{d}= r\sin\varphi_{1}\sin\varphi_{2}...\sin\varphi_{d-2}\sin\varphi_{d-1}$, the polynomials take the form $P_{l,d}(\textbf{r}) = r^{l}Y_{l,d}(\theta)$, where $Y_{l,d}$ is a $d$-dimensional spherical harmonic of degree $l$ and $\theta$ refers to the point on unit sphere with angular coordinates $\varphi_{1}$, $\varphi_{2}$,...,$\varphi_{d-1}$.  For $l\neq 0$ there exist
\begin{equation}
\label{degeneracy}
m_{l,d}= \frac{(2l+d-2)(d+l-3)!}{(d-2)!l!}
\end{equation}
linearly-independent spherical harmonics of order $l$.  Following Mikhlin \cite{Mikhlin}, these will be denoted by $Y_{l,d}^{(m)}(\theta)$, $m = 1,2, ...,m_{l,d}$.  For any $l$ the functions $Y_{l,d}^{(m)}(\theta)$ can be made orthonormal on the unit sphere, and we assume this is the case.

The remaining steps mirror our discussion of the two- and three-dimensional cases.  The boundary-value problem (\ref{Helmholtz}) that has to be solved is 
\begin{equation}
\label{bvdsphere}
\left (\frac{\partial^{2}}{\partial r^{2}} +\frac{d-1}{r}\frac{\partial}{\partial r} + \frac{1}{r^{2}}\triangle_{\theta} - \frac{\omega^{2}}{c^{2}}\right )u_{\omega} = 0, ~~u_{\omega}|_{a} = f_{\omega}(\theta)
\end{equation}
where $\triangle_{\theta}$ stands for a multiplicative angular piece of the Laplacian;  explicit expressions for $\triangle_{\theta}$ in two and three dimensions are given in Eqs.(\ref{bvcircle}) and (\ref{bvsphere}), respectively.   Seeking the particular solution in the form, $u_{\omega}(r, \theta) = R_{\omega}(r) Y_{l,d}^{(m)}(\theta)$, we find that the radial function $R_{\omega}(r)$ satisfies the equation
\begin{equation}
\label{dspBessel}
\frac{d^{2}R_{\omega}}{dr^{2}} + \frac{d-1}{r}\frac{dR_{\omega}}{dr} - \left (\frac{\omega^{2}}{c^{2}} + \frac{l(l+d-2)}{r^{2}}\right )R_{\omega} = 0
\end{equation}
where we employed the result $\triangle_{\theta}Y_{l,d}^{(m)}(\theta) = - l(l+d-2)Y_{l,d}^{(m)}(\theta)$ \cite{Sommerfeld,Muller}.  Linearly-independent solutions to (\ref{dspBessel}) are $r^{-\frac{d-2}{2}}I_{l + \frac{d-2}{2}}(|\omega|r/c)$ and $r^{-\frac{d-2}{2}}K_{l + \frac{d-2}{2}}(|\omega|r/c)$ \cite{Sommnote} thus implying that the particular solution to the boundary-value problem (\ref{bvdsphere}) finite at the origin $r = 0$ and decaying as $r \rightarrow \infty$ is 
\begin{eqnarray}
\label{dsphere solution}
u_{\omega}& = &\sum_{l=0}^{\infty}\sum_{m=1}^{m_{l,d}}\frac{a^{\frac{d-2}{2}}I_{l+\frac{d-2}{2}}(|\omega |r/c)}{r^{\frac{d-2}{2}}I_{l+\frac{d-2}{2}}(|\omega |a/c)} f_{\omega l}^{(m)} Y_{l,d}^{(m)}(\theta), ~r \leqslant a \nonumber\\
u_{\omega}& = &\sum_{l=0}^{\infty}\sum_{m=1}^{m_{l,d}}\frac{a^{\frac{d-2}{2}}K_{l+\frac{d-2}{2}}(|\omega |r/c)}{r^{\frac{d-2}{2}}K_{l+\frac{d-2}{2}}(|\omega |a/c)} f_{\omega l}^{(m)} Y_{l,d}^{(m)}(\theta), ~r > a \nonumber\\
\end{eqnarray}
At the sphere surface these reduce to $f_{\omega}(\theta)= \sum_{l,m}f_{\omega l}^{(m)}Y_{l,d}^{(m)}(\theta)$ which is a series expansion in spherical harmonics of an arbitrary function defined on a sphere \cite{Muller, Mikhlin}.  Substituting the solution (\ref{dsphere solution}) in Eq.(\ref{actionsurfaceintegral}) and performing the integration we find
\begin{equation}
\label{dsphereaction}
S_{E} = \frac{\hbar a^{d-2}}{2T}\sum_{\omega,l,m}\frac{|f_{\omega l}^{(m)}|^{2}}{I_{l+\frac{d-2}{2}}(|\omega|a/c)K_{l+\frac{d-2}{2}}(|\omega|a/c)}
\end{equation}
As a result the Casimir energy $\mathcal{E}$ and the pressure on the boundary $\mathcal{P} = -(S_{d}a^{d-1})^{-1}\partial \mathcal{E}/\partial a$  (here $S_{d}=2\pi^{d/2}/\Gamma(d/2)$ is the surface area of a unit sphere) are given by
\begin{equation}
\label{Casenergydsphere}
\mathcal{E} = \frac{\hbar c}{2\pi}\sum_{l=0}^{\infty}m_{l,d}\int_{0}^{\infty}d\kappa
\ln \frac{I_{l+\frac{d-2}{2}}(\kappa a) K_{l+\frac{d-2}{2}}(\kappa a)}{(\kappa a)^{d-2}}
\end{equation} 
\begin{eqnarray}
\label{Caspressuredsphere}
\mathcal{P} &= &- \frac{\hbar c}{2\pi S_{d}a^{d+1}}\sum_{l=0}^{\infty}m_{l,d}\nonumber\\&\times&\int_{0}^{\infty}xdx\frac{d}{dx}\ln \frac{I_{l+\frac{d-2}{2}}(x)K_{l+\frac{d-2}{2}}(x)}{x^{d-2}}
\end{eqnarray}
which are the results due to Bender and Milton \cite{Bender};  specifically the expression for the pressure  is identical to their Eq.(3.5).   

In three dimensions Eqs.(\ref{Casenergydsphere}) and (\ref{Caspressuredsphere}) clearly reduce to Eqs. (\ref{Casenergysphere}) and (\ref{Caspressuresphere}).  In order to see that the two-dimensional results are  also reproduced, we recall that \cite{Mikhlin} for $d=2$ and $l >0$ there are  only two linearly-independent harmonic polynomials of degree $l$, namely the real and imaginary parts of $(x_{1} + ix_{2})^{l}$, while for $d=2$ and $l=0$ there is only one polynomial which is a constant.  This implies that in two dimensions the degeneracy factor (\ref{degeneracy}) satisfies the rules, $m_{l,2} = 2$ for $l>0$ and $m_{0,2}=1$.  With this in mind we see that for $d=2$ Eqs.(\ref{Casenergydsphere}) and (\ref{Caspressuredsphere}) reproduce Eqs.(\ref{Casenergycircle}) and (\ref{Casforcecircle}).

\section{Divergences and their geometrical interpretation}

Although Eq.(\ref{planarpressure}) gives a finite universal Casimir pressure for the planar version of the problem,  its circular (\ref{Casforcecircle}), cylindrical (\ref{Caspressurecylinder}), spherical (\ref{Caspressuresphere}), (\ref{Caspressuredsphere}) (and any curvilinear) counterparts are divergent.  Various techniques have been successfully used to remove the divergences in the case of a cylinder \cite{Nesterenko} and in spherical \cite{Bender} geometries of odd dimensionality predicting universal Casimir pressures.  However no consistent removal procedure was found capable of handling the spherical geometry of even space dimension \cite{Bender}.  This includes the experimentally relevant two-dimensional case where as an alternative, Sen \cite{Sen} proposed to view the action (\ref{action}) as an effective low-energy theory to be supplemented by a cutoff function like in Eq.(\ref{Casenergygeneral}).  This removes the divergence, and leads to a finite non-universal effect.  

In order to understand the difference between the planar and curved geometries we notice that the Casimir force is the change of the energy upon infinitesimal displacement of the boundary.  A geometrically sharp boundary possesses a divergent energy per unit area, coming from the exclusion of the high frequency and short wavelength modes from an increasingly narrow region near the surface.  This divergent self-energy does not contribute to the Casimir force in the planar case because the overall area remains fixed as the boundary is displaced.  However this is not the case for curved boundaries.  Indeed a change of the radius of a circle implies a change of the perimeter and as a result the divergent self-energy will contribute into the force.   This idea originally due to Deutsch and Candelas\cite{CD} was recently re-expressed by Graham, Jaffe and co-workers \cite{Jaffe} and by Barton \cite{Barton}.  The implication is that a real-world curved boundary may be responsible for large non-universal portion of the physically measurable Casimir force \cite{critique}.  This is consistent with Sen's observation \cite{Sen} that the coefficients of the non-universal terms contributing into the Casimir energy are geometrical objects such as the length of the circular boundary, etc.  We also note that in the high-temperature limit the Casimir \textit{free energy} can be shown to have entirely geometrical nature - it is expressed in terms of a surface integral of a quadratic function of local curvature \cite{BD}.     

The geometry-dependent contributions to the Casimir effect can be classified into three groups:  (i)  non-universal terms, which diverge as a power of the cutoff frequency $\omega_{0}$;  (ii) nearly-universal terms, which diverge as a logarithm of $\omega_{0}$; and (iii) universal terms, which remain finite as $\omega_{0} \rightarrow \infty$.  The nearly-universal contributions are characterized by universal amplitudes and a weak dependence on the choice of the cutoff function. 

A further test of the idea of Refs.\cite{CD,Jaffe,Barton} would consist in calculation of the effect in a curved geometry where various contributions into the Casimir interaction can be unambiguously separated.  A natural candidate is a system characterized by more than one macroscopic length scale where the area or circumference of the boundary is independent of its curvature.  
\begin{figure}
\includegraphics[width=1.0\columnwidth,keepaspectratio]{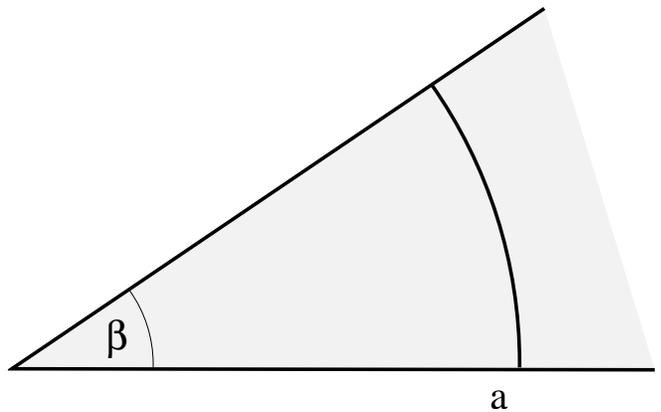} 
\caption{Wedge of opening angle $\beta$ with superimposed arc of radius $a$ in two dimensions.}
\end{figure}

\subsection{Wedge-circular arc geometry in two dimensions;  Dirichlet case}

Consider a wedge of opening angle $\beta$ with superimposed arc of radius $a$, Fig. 1.  This is a geometry in which both a Casimir force and torque exist.

The semi-circular ($\beta =\pi$) version of this problem has been studied earlier \cite{Nesterenko2}  using the zeta-function regularization technique.  Similar to the circular case, a divergence was found which led the authors to conclude that "for obtaining the physical result an additional renormalization is needed".  

The wedge-arc configuration with an arbitrary angle $\beta$ has been
considered in Ref. \cite{Nesterenko3} with the aim of revealing the
regularities in the boundary non-smoothness contributions  to
the heat kernel coefficients. The local characteristics of the vacuum
have been discussed by Sakharian and collaborators for a scalar field with Dirichlet boundary
condition in general space-time dimension \cite{Sakharian1} and for the electromagnetic field in
three-dimensional space \cite{Sakharian2}. 

Although the Casimir interaction can be obtained from the Casimir energy density of Ref.\cite{Sakharian1}, our approach provides a quicker route to the final result.  Indeed the  Casimir energy can be inferred from the results for the circular geometry since the boundary-value problem we need to solve is closely related to (\ref{bvcircle}).  The  difference is that we seek a solution inside the Dirichlet wedge $0 \leqslant \varphi \leqslant \beta$, thus implying $u_{\omega}(\rho,\varphi) = R_{\omega}(\rho) \sin(\pi n \varphi/\beta)$, $n = 1, 2, ...$ for the particular solution. The radial function $R_{\omega}(\rho)$ satisfies the same Eq.(\ref{Bessel}) with $n$ being  replaced by $\pi n/\beta$.  As a result the solution in question can be obtained from Eq.(\ref{circle solution}) by replacing the order $n$ of the Bessel functions with $\pi n/\beta$, the angular function $\exp i n\varphi$ with $\sin(\pi n\varphi/\beta)$, and restricting the summation over $n$ from unity to infinity.  The calculation of the Casimir energy is similar to that for the circular geometry with the result
\begin{eqnarray}
\label{casenergywedge}
&\mathcal{E}& = \frac{\hbar c}{2\pi a}\sum_{n = 1}^{\infty} \int_{0}^{\infty} dx \ln \frac{I_{\frac{\pi n}{\beta}}(x)K_{\frac{\pi n}{\beta}}(x)}{\beta}\nonumber\\
&\rightarrow& -\frac{\hbar c}{2s}\sum_{n=1}^{\infty}\int_{0}^{\infty} ntdtF(\frac{\pi nct}{\omega_{0}s})\frac{d}{dt}\ln\frac{I_{\frac{\pi n}{\beta}}(\frac{\pi n}{\beta}t)K_{\frac{\pi n}{\beta}}(\frac{\pi n}{\beta}t)}{\beta}\nonumber\\
\end{eqnarray} 
where in the second representation we restored the cutoff function (see Eq.(\ref{Casenergygeneral})) and introduced the length of the arc $s=\beta a$.
We then employ the uniform asymptotic $n \gg 1$ expansion of Debye \cite{AS,Schwinger}:
\begin{eqnarray}
\label{Debye}
2n(1&+&y^{2})^{1/2}I_{n}(ny)K_{n}(ny) = 1 + \frac{1}{8n^{2}} \Big\{\frac{1}{1+y^{2}}\nonumber\\& -& \frac{6}{(1+y^{2})^{2}} + \frac{5}{(1+y^{2})^{3}} \Big\} + \mathcal{O}(\frac{1}{n^{3}})
\end{eqnarray}  
which can be used to evaluate the energy (\ref{casenergywedge}) in the $\beta \ll 1$ limit.  To leading order we find
\begin{eqnarray}
\label{firstorder}
\mathcal{E}^{(1)}& =&  \frac{\hbar c}{2s}\sum_{n=1}^{\infty}\int_{0}^{\infty}ntdtF(\frac{\pi nct}{\omega_{0}s})\frac{d}{dt}\ln \left (2\pi n(1+t^{2})^{1/2}\right )\nonumber\\
& = &\frac{\hbar c}{4\pi s}\int_{0}^{\infty}xdxF(\frac{cx}{\omega_{0}s})\sum_{n=1}^{\infty}\frac{2x}{x^{2}+\pi^{2}n^{2}}\nonumber\\
&=& \frac{\hbar c}{4\pi s}\int_{0}^{\infty}dxF(\frac{cx}{\omega_{0}s})\left (x(\coth x-1) -1 +x \right )
\end{eqnarray}
where in going from the second to the third representation we used Euler's sum
\begin{equation}
\label{Euler}
\sum_{n=1}^{\infty}\frac{1}{x^{2}+(\pi n)^{2}}= \frac{1}{2x}\left (\coth x -\frac{1}{x}\right )
\end{equation}
Although the expression (\ref{firstorder}) is divergent in the $\omega_{0} \rightarrow \infty$ limit, it is important to realize that it depends on $a$ and $\beta$ only through the arc length $s = \beta a$.  In order to interpret (\ref{firstorder}) geometrically we note that the dimensionless combination $\omega_{0}s/c$ entering the argument of the cutoff function is a ratio of the macroscopic $s$ and microscopic $c/\omega_{0}$ length scales of the problem, and naturally one has $s \gg c/\omega_{0}$ which is the range of applicability of the theory.  Then in order to evaluate the first term of (\ref{firstorder}) we expand the cutoff function in a Taylor series around zero.  The leading $F(0) = 1$ term of the expansion then gives an integral of the form (\ref{integral}), thus generating a universal $\pi \hbar c/48s$ contribution to the Casimir energy.  All higher order contributions of the expansion are convergent and proportional to negative powers of the cutoff frequency - in the $\omega_{0}s/c \gg 1$ limit they all vanish.   The second term of Eq.(\ref{firstorder}) can be ignored because it leads to a finite geometry-independent energy contribution of the order $\hbar \omega_{0}$, while the contribution from the last term is linear in the arc length.   As a result we arrive at 
\begin{equation}
\label{finalfirstorder}
\mathcal{E}^{(1)}= \alpha s + \frac{\pi \hbar c}{48s}, ~~~\alpha=\frac{\hbar \omega_{0}^{2}}{4\pi c}\int_{0}^{\infty}tF(t)dt
\end{equation}
where $\alpha$ is the line tension coefficient;  as expected, it is of the order of the microscopic energy scale $\hbar\omega_{0}$ divided by the microscopic length scale $c/\omega_{0}$.  If we specialize to the case of an exponential cutoff function ($F(t) = \exp(-t)$), then our expression for the line tension reproduces Sen's result $\alpha = \hbar \omega_{0}^{2}/(4 \pi c)$ \cite{Sen}.

Even though Eq.(\ref{finalfirstorder}) is the leading order term of the $\beta \ll 1$ expansion of the Casimir energy, there is no restriction on the segment length $s = \beta a$ since the radius $a$ can be arbitrary.  We see that to the leading order there is only the dependence on the segment length (thus implying locality) but not on its curvature.  Eq.(\ref{finalfirstorder}) also predicts that if the system were given the freedom to choose the optimal value of $s$ by minimizing the energy, this would be  a microscopic  $s \sim c/\omega_{0}$ which is a compromise between the energy decrease due to shrinking of the segment length (the non-universal term) and repulsive universal piece of the Casimir interaction preferring large segment lengths.  The precise value of the optimal $s$ implied by (\ref{finalfirstorder}) is not to be trusted as $s \sim c/\omega_{0}$ is at the verge of applicability of the theory.   If we take the simultaneous $\beta \rightarrow 0$, $a\rightarrow \infty$ limit but keep the length $s = \beta a$ fixed, the geometry of Fig.1 turns into that of an infinite strip of width $s$ whose sides are connected by a straight Dirichlet bridge;  its Casimir energy is given by Eq.(\ref{finalfirstorder}).  From this viewpoint the universal $\pi\hbar c/48s$ piece of (\ref{finalfirstorder})  can be understood by approximating the bridge by a strictly one-dimensional Dirichlet interval.  The Casimir energy of a free field confined to an interval of length $s$ is $-\pi\hbar c/24s$ and attractive \cite{Boyer}.  In our case it is repulsive as these fluctuations are eliminated by the boundary.  Additionally, the magnitude of our effect is smaller since the one-dimensional approximation incorrectly assumes an infinitely sharp field localization and thus a larger change of the vacuum energy.

In order to improve on Eq.(\ref{finalfirstorder}) we subtract $\mathcal{E}^{(1)}$, Eq.(\ref{firstorder}), from Eq.(\ref{casenergywedge}).  The outcome $\mathcal{U} = \mathcal{E} - \mathcal{E}^{(1)}$ is expected to be universal or nearly-universal since the largest non-universal fraction of the effect is already included in $\mathcal{E}^{(1)}$:  
\begin{equation}
\label{wedgeenergyuniversal}
\mathcal{U} = \frac{\hbar c}{2s}\sum_{n=1}^{\infty}n\int_{0}^{\infty} dt \ln(2\frac{\pi n}{\beta}(1+t^{2})^{\frac{1}{2}}I_{\frac{\pi n}{\beta}}(\frac{\pi n}{\beta}t) K_{\frac{\pi n}{\beta}}(\frac{\pi n}{\beta}t))
\end{equation}
where we suppressed the cutoff function.
Using the expansion (\ref{Debye}) again we evaluate the energy (\ref{wedgeenergyuniversal})  to the next order in $\beta \ll 1$.  If the integration is performed first, then we find
\begin{equation}   
\label{wedgeenergydivergent}
\mathcal{U} = - \frac{\hbar c \beta^{2}}{256\pi s}\sum_{n=1}^{\infty}n^{-1}, ~~~\beta \ll 1, 
\end{equation}
which is marginally divergent.  This means that setting the cutoff function at unity is not justified, and that there is a weak dependence on the cutoff frequency $\omega_{0}$.  Therefore we restore the cutoff function, go beyond the leading order term in the Debye expansion (\ref{Debye}) and write Eq.(\ref{wedgeenergyuniversal}) as
\begin{eqnarray}
\label{alt1}
\mathcal{U} &= &-\frac{\hbar c\beta^{2}}{16\pi s}\int_{0}^{\infty}xdxF(\frac{cx}{\omega_{0}s})\frac{d}{dx}\sum_{n=1}^{\infty}\Big\{\frac{1}{x^{2}+(\pi n)^{2}}\nonumber\\
&-&\frac{6(\pi n)^{2}}{(x^{2} + (\pi n)^{2})^{2}} + \frac{5(\pi n)^{4}}{(x^{2}+(\pi n)^{2})^{3}}\Big\}
\end{eqnarray}
The sums over $n$ can be computed with the help of Euler's formula (\ref{Euler}) and two consequent relationships:
\begin{equation}
\label{sum2}
\sum_{n=1}^{\infty}\frac{(\pi n)^{2}}{(x^{2}+(\pi n)^{2})^{2}} = \frac{(x \coth x)'}{4x}
\end{equation}
\begin{equation}
\label{sum3}
\sum_{n=1}^{\infty}\frac{(\pi n)^{4}}{(x^{2}+(\pi n)^{2})^{3}} = \frac{(x^{3}(x \coth x)')'}{16x^{3}}
\end{equation}
where the prime stands for the derivative with respect to $x$.   Comparing the right- and left-hand sides of Eqs.(\ref{Euler}), (\ref{sum2}), and (\ref{sum3}) we observe that the sum $\sum \Big\{...\Big\}$ in Eq.(\ref{alt1}) approaches a constant limit as $x\rightarrow 0$.  This implies that the integral over $x$ is convergent at the lower limit.  On the other hand, in the $x\rightarrow \infty$ limit we find that  $\sum \Big\{...\Big\} \rightarrow -1/(2^{7}x)$ with the implication that if the cutoff function is set to unity, then, consistent with Eq.(\ref{wedgeenergydivergent}), the integral over $x$ would logarithmically diverge at the upper limit.  The role of the cutoff function consists in restoring the convergence by effectively setting the upper limit of the integration at $\omega_{0}s/c \gg 1$.  In view of the logarithmic character of the divergence, the outcome is rather insensitive to the  upper integration limit.  As a result we arrive at the nearly-universal contribution to the Casimir energy
\begin{equation}
\label{final}
\mathcal{U} = - \frac{\hbar c \beta^{2}}{256\pi s}\ln\frac{\omega_{0}s}{c}, ~~~\beta \ll 1
\end{equation} 
which has logarithmic accuracy.

The result (\ref{final}) can be also understood heuristically starting from the divergent expression (\ref{wedgeenergydivergent}) and employing the equality $\sum_{n=1}^{N}n^{-1} = C + \ln N + \epsilon_{N}$ where $C$ is Euler's constant and $\epsilon_{N} \rightarrow 0$ as $N \rightarrow \infty$ \cite{AS}.  The parameter $N$ is estimated by recalling that the suppressed cutoff function $F(\pi nct/\omega_{0}s)$ effectively ends the sum of $1/n$ at $n = N$ such as $\pi nct/\omega_{0}s \approx 1$ and the relevant value of $t$, as implied by the second-order term of the Debye expansion (\ref{Debye}), is of the order unity.  Thus $N \approx \omega_{0}s/c \gg 1$ and we recover Eq.(\ref{final}).

The result (\ref{final}) implies that for \textit{fixed} arc length $s$ when the leading order contribution (\ref{finalfirstorder}) can be regarded as a constant, the zero-point motion induces a \textit{widening} torque $-\partial \mathcal{U}/\partial \beta$ which could be expected on physical grounds.  Combining Eqs.(\ref{finalfirstorder}) and (\ref{final}) we arrive at the expression for the Casimir energy of the Dirichlet wedge of arc length $s$ and opening angle $\beta= s/a \ll 1$ 
\begin{equation}
\label{totenergy}
\mathcal{E} = \alpha s- \frac{\hbar cs}{256\pi a^{2}}\ln\frac{\omega_{0}s}{c}+ \frac{\pi \hbar c}{48s}
\end{equation}
This result supports a geometrical interpretation for the non-universal contributions to the Casimir energy.  Indeed, when the frequency cutoff $\omega_{0}$ tends to infinity, the strongest ($\omega_{0}^{2}$) divergence comes from the first term of (\ref{totenergy}) proportional to the length of the boundary thus supporting the idea of Refs.\cite{CD,Jaffe,Barton}.   The sub-leading logarithmically divergent term of (\ref{totenergy}) can be also understood geometrically as being proportional both to the length of the boundary and to the square of its curvature.  This term can be alternatively interpreted as a finite-size correction to the surface tension: $\alpha \rightarrow \alpha - \hbar c\ln(\omega_{0}s/c)/256\pi a^{2}$.  We note that Eq.(\ref{totenergy}) has a local character and exhibits a term-by-term correspondence with its circular counterpart \cite{Sen}.      

The accuracy of the Debye expansion (\ref{Debye}), implies that Eq.(\ref{totenergy}) approximately captures the whole $0\leqslant \beta \leqslant 2\pi$ range.  

\subsection{Wedge-circular arc geometry in two dimensions:  periodic boundary conditions}

The Casimir effect is sensitive to changes in space topology.  In our next example we demonstrate that the non-universal contributions to the Casimir effect do not seem to be sensitive to topology while the universal ones are.  Consider the wedge-circular arc geometry as in Fig. 1 but assume that the wedge edges are physically identical.  This is implemented by imposing the condition of periodicity,  $u(\rho, \varphi) = u(\rho, \varphi+\beta)$, where the opening angle $\beta$ can now be arbitrary.  This means that the field $u$ belongs to a conical surface with a Dirichlet circle which is a distance $a$ away from cone's apex as measured along the cone surface \cite{Nesterenko3}.  For $\beta=2\pi$ the geometry becomes that of a plane with the Dirichlet circle. 

The Casimir energy of this configuration can be inferred from the solutions of the circular and wedge cases.  Indeed, the boundary-value problem we need to solve is posed by Eq.(\ref{bvcircle}) with the boundary field satisfying the condition of periodicity, $f_{\omega}(\varphi) = f_{\omega}(\varphi+\beta)$.  This implies $u_{\omega}(\rho,\varphi) = R_{\omega}(\rho) \sin(2\pi n \varphi/\beta)$, $n = 1, 2, ...$ or $u_{\omega}(\rho,\varphi) = R_{\omega}(\rho) \cos(2\pi n \varphi/\beta)$, $n = 0, 1, 2, ...$ for the particular solution. In both cases the radial function $R_{\omega}(\rho)$ satisfies the same Eq.(\ref{Bessel}) with $n$ being  replaced by $2\pi n/\beta$.  The Casimir energy can then be inferred from Eq.(\ref{casenergywedge}) as  
\begin{eqnarray}
\label{Casenergyperiodicwedge}
\mathcal{E}& =& \frac{\hbar c}{2\pi a}\int_{0}^{\infty} dx \ln \frac{I_{0}(x)K_{0}(x)}{\beta}\nonumber\\
&+& \frac{\hbar c}{\pi a}\sum_{n = 1}^{\infty} \int_{0}^{\infty} dx \ln \frac{I_{\frac{2\pi n}{\beta}}(x)K_{\frac{2\pi n}{\beta}}(x)}{\beta}
\end{eqnarray}
where the first term is due to the fluctuating angle-independent Fourier component of the boundary field $f_{\omega}$ while the sum is a contribution from the angle-dependent components.  The latter are represented by a sine and a cosine Fourier series contributing equally into the Casimir energy.  We note that for $\beta=2\pi$ Eq.(\ref{Casenergyperiodicwedge}) reduces to its circular counterpart, Eq.(\ref{Casenergycircle}).   

In order to understand the geometry dependence of the first term of (\ref{Casenergyperiodicwedge}) we restore the cutoff function $F(y)$:
\begin{equation}
\label{n=0term}
\mathcal{E}_{0} = -\frac{\hbar c}{2\pi a}\int_{0}^{\infty} xdx F(\frac{cx}{\omega_{0}a})\frac{d}{dx}\ln\frac{I_{0}(x)K_{0}(x)}{\beta}
\end{equation}
The derivative removes the dependence on the opening angle $\beta$.  
 Since  $2xI_{0}(x)K_{0}(x) \approx 1$ \cite{AS} for large values of the
argument, the integral (\ref{n=0term}) diverges linearly at the upper limit
when the cutoff function is set at unity.  The presence of the cutoff function
effectively sets the upper integration limit  at $\omega_{0}a/c$ thus 
leading to a geometry-independent estimate $\mathcal{E}_{0} \sim \hbar \omega_{0}$. 
The $
 a$-dependent part can be found by subtracting from (\ref{n=0term}) the 
 same integral with $I_{0}(x)K_{0}(x)$ replaced with its large argument limit $1/2x$:
\begin{equation}
\label{adependentuniversal}
\mathcal{E}_{0} \rightarrow \frac{\hbar c}{2\pi a}\int_{0}^{\infty} dx \ln 2xI_{0}(x)K_{0}(x) = - \frac{0.08808\hbar c}{2\pi a} 
\end{equation} 
This is the universal piece of the Casimir effect when the cutoff function is set at unity (i. e. the $\omega_{0}a/c \rightarrow \infty$ limit is taken). The integral was numerically evaluated by Sen \cite{Sen} and more accurately by Milton and Ng \cite{Ng} whose value is used in (\ref{adependentuniversal}).

The second term of Eq. (\ref{Casenergyperiodicwedge}) can be analyzed in a manner similar to that of Eq.(\ref{casenergywedge}).  The results of such analysis combined with (\ref{adependentuniversal}) give the Casmir energy for the wedge-arc geometry with periodic boundary conditions:
\begin{equation}
\label{totalenergyperiodicwedge}
\mathcal{E} = \alpha s- \frac{\hbar cs}{256\pi a^{2}}\ln\frac{\omega_{0}s}{c}+ \frac{\pi \hbar c}{12s}- \frac{0.08808\hbar c}{2\pi a} 
\end{equation}
where the coefficient of surface tension is defined in Eq.(\ref{finalfirstorder}).   We see that the cutoff-dependent terms of Eqs.(\ref{totenergy}) and (\ref{totalenergyperiodicwedge}) are the same.  At the same time the effect of topological change in going from the Dirichlet to periodic boundary conditions manifests itself in the universal parts of the Casimir effect.  Indeed, the $s$-dependent piece of the interaction, $\pi\hbar c/12s$, is four times larger than its Dirichlet counterpart, $\pi\hbar c/48s$, in Eqs.(\ref{finalfirstorder}) and (\ref{totenergy}).  This mirrors the relationship between the Dirichlet and periodic Casimir energies for the free field in one dimension \cite{conformal}.  Additionally, the last attractive term of (\ref{totalenergyperiodicwedge}) is unique to the periodic geometry.

For the special case of a circle in a plane, $s = 2\pi a$, Eq. (\ref{totalenergyperiodicwedge}) simplifies to 
\begin{equation}
\label{Sencircle}
\mathcal{E} = 2\pi\alpha a - \frac{\hbar c}{128a}\ln\frac{\omega_{0}a}{c}+\frac{0.055\hbar c}{2a}
\end{equation}
This agrees with Sen's result \cite{Sen} except for the magnitude of the last
term.  We believe Sen's value 0.045 is a typographical error (according to the
numerical results he quotes, the value should be 0.054).

Similar to the argument following Eq.(\ref{finalfirstorder}) we can take the simultaneous $\beta \rightarrow 0$, $a\rightarrow \infty$ limit but keep the segment length $s = \beta a$ fixed.  Then Eq.(\ref{totalenergyperiodicwedge}) simplifies to 
\begin{equation}
\label{cylinderneck}
\mathcal{E}^{(1)}= \alpha s + \frac{\pi \hbar c}{12s}
\end{equation}  
which is a periodic counterpart of Eq.(\ref{finalfirstorder});  it gives the Casimir energy of a Gaussian field defined on an infinite cylinder of radius $s/2\pi$ in the presence of a Dirichlet circle belonging to the cylinder surface.

\section{Arbitrary smooth boundary of a compact region}

Since for the circular geometry the length of the boundary and its curvature are not independent, a route different from direct calculation is needed to demonstrate the geometrical nature of the non-universal contributions to the Casimir interaction, Eq.(\ref{Sencircle}).  For that purpose Sen \cite{Sen} employed a result due to Pleijel  \cite{geometry} which relates asymptotic behavior of a certain integral of a Green's function to geometry.  In the process Sen derived a remarkable expression for the non-universal part of the Casimir energy $\mathcal{E}(\omega_{0})$ for an arbitrary smooth boundary of a compact region.  Here we show how this result can be understood heuristically by employing our findings for the wedge geometry. 

First we notice that for the wedge whose edges are Dirichlet/periodically connected by an infinitesimally small arc/circle of length $ds$ the non-unversal part of the Casimir energy (see Eqs. (\ref{wedgeenergydivergent}), (\ref{totenergy}) and (\ref{totalenergyperiodicwedge})) can be rewritten as a differential relationship
\begin{equation}
\label{differential}
d\mathcal{E}(\omega_{0}) = \alpha ds + d\mathcal{U} = \alpha ds - \frac{\hbar c}{256\pi}\mathcal{C}^{2}(s)ds\sum_{n=1}^{\infty}n^{-1}
\end{equation}
where $\mathcal{C} = 1/a$ is the curvature of the arc.  Our calculations indicate that the energy (\ref{differential}) has to be attributed to the arc itself rather than to the conditions imposed on the wedge edges;  those conditions determine the universal part of the Casimir effect.  With this in mind let us consider an arbitrary smooth boundary of a compact region.  Then Eq.(\ref{differential}) can be interpreted as representing a non-universal contribution into the Casimir energy due to the infinitesimally small boundary element $ds$ whose position, the length $s$ along the boundary, is measured relative to an arbitrary reference point and $\mathcal{C}(s)$ is the curvature at location $s$.

For an arbitrary smooth boundary the non-universal piece of the Casimir energy can be found by integrating (\ref{differential}) over the length of the boundary
\begin{equation}
\label{integraldivergent}
\mathcal{E}(\omega_{0}) = \alpha S - \left (\frac{\hbar c}{256\pi}\int\mathcal{C}^{2}(s)ds\right )\sum_{n=1}^{\infty}n^{-1}
\end{equation}
where $S= \int ds$ is the total length of the boundary.  We notice that the marginally divergent sum of $n^{-1}$ does not have any reference to specific geometry.  This  leads us to a \textit{conjecture} that (similar to the case of the circular arc) the effect of the frequency cutoff $\omega_{0}$  consists in replacement of the divergent sum in (\ref{integraldivergent}) with the finite logarithm of the large ratio of the macroscopic and microscopic length scales:
\begin{equation}
\label{integralfinite}
\mathcal{E}(\omega_{0}) = \alpha S - \left (\frac{\hbar c}{256\pi}\int\mathcal{C}^{2}(s)ds\right )\ln\frac{\omega_{0}S}{c}
\end{equation}
Since Eq.(\ref{integralfinite}) has logarithmic accuracy, the precise meaning of the macroscopic length scale is not essential - we have chosen it to be the total length $S$ of the boundary.   The result (\ref{integralfinite}) is due to Sen \cite{Sen,typo}.  Our heuristic argument which leads to Sen's result (\ref{integralfinite}) also links the Pleijel formula \cite{geometry} to the Debye expansion (\ref{Debye}) which is in the heart of our wedge results.    

The remarkable feature of Eq.(\ref{integralfinite}) is that it is solely determined by local geometry of the boundary.  In arriving at (\ref{integralfinite}) we employed additivity of the interaction which is known generally not to be the case for Casimir interactions.  We note however that the additivity only holds for the strictly non-universal piece of the effect.  For the nearly-universal second term of Eq.(\ref{integralfinite}) the additivity is logarithmically weakly violated.  This allowed us to proceed by assuming the additivity and to deal with the logarithmic divergence separately.  Additionally, our calculations indicate that additivity does not hold for the universal parts of the Casimir energy which could not be expressed in a differential form.  The implication is that it would be very difficult if not impossible to come up with a purely geometrical formula for the universal part of the Casimir effect.  

\section{Summary}

To summarize, we have demonstrated how Casimir effects caused by sharp boundaries can be efficiently computed by focusing on the quantum fluctuations eliminated by these boundaries.  The applicability of this method is not limited to the scalar field theory (\ref{action}), Dirichlet boundaries, zero-temperature limit or to the geometries we have considered.  

Second, we presented an explicit calculation to support the idea \cite{CD,Jaffe,Barton} that the divergent Casimir forces encountered in the presence of curved boundaries have geometrical origin - they are largely due to divergent self-energy contributions.  Our analysis also supports Sen's  proposal \cite{Sen} that the Casimir effect in a Dirichlet ring in two dimensions is finite and non-universal with the cutoff frequency $\omega_{0}$ supplied by the properties of the material the boundary is made of.  More work is needed to further explore the geometrical nature of the non-universal parts of the Casimir effect as they may be responsible for largest contributions into experimentally measurable Casimir force. 

\section{Acknowledgments}

We thank A. A. Saharian for informing us of his work on a related topic.  This work was supported by the Thomas F. Jeffress and Kate Miller Jeffress Memorial Trust.


\begin{thebibliography}{10}
\bibitem{QMnote}  An equation which is only different from (\ref{spBessel}) in the sign of the $\omega^{2}/c^{2}$ combination is solved in Section 33 of Ref. \cite{QM}.
\bibitem{Sommnote}  An equation which is only different from (\ref{dspBessel}) in the sign of the $\omega^{2}/c^{2}$ combination is solved in Section B of Ref. \cite{Sommerfeld}.
\bibitem{typo} Sen's result \cite{Sen} is inadvertantly missing a factor of $\pi$ in the 
denominator of the second term.
\end{thebibliography}
\end{document}